# Predator-Prey Interactions between Droplets Driven by Nonreciprocal Oil Exchange

**Authors:** Caleb H. Meredith[1†], Pepijn G. Moerman[2,3†], Jan Groenewold[2,4], Yu-Jen Chiu[1], Willem K. Kegel[2], Alfons van Blaaderen[3], Lauren D. Zarzar[1,5,6]*

**Affiliations:**
1. Department of Materials Science and Engineering, The Pennsylvania State University, PA 16802, USA
2. Van't Hoff Laboratory for Physical & Colloid Chemistry, Debye Institute for Nanomaterials Science, Utrecht University, The Netherlands
3. Soft Condensed Matter, Debye Institute for Nanomaterials Science, Utrecht University, The Netherlands
4. Guangdong Provincial Key Laboratory of Optical Information Materials and Technology & Institute of Electronic Paper Displays, South China Academy of Advanced Optoelectronics, South China Normal University, Guangzhou 510006, P. R. China.
5. Department of Chemistry, The Pennsylvania State University, PA 16802, USA
6. Materials Research Institute, The Pennsylvania State University, PA 16802, USA

† These authors contributed equally to this work
*Correspondence to: ldz4@psu.edu

**Abstract**: Chemotactic interactions are ubiquitous in nature and can lead to nonreciprocal and complex emergent behavior in multibody systems. Here we show how chemotactic signaling between microscale oil droplets of different chemistries in micellar surfactant solutions can result in predator-prey-like chasing interactions. The interactions and dynamic self-organization result from the net directional, micelle-mediated transport of oil between emulsion droplets of differing composition and are powered by the free energy of mixing. The nonreciprocal behavior occurs in a wide variety of oil and surfactant conditions, and we systematically elucidate chemical design rules for tuning the interactions between droplets by varying oil and surfactant chemical structure and concentration. Through integration of experiment and simulation, we also investigate the active behavior and dynamic reorganization of multi-droplet clusters. Our findings demonstrate how chemically-minimal systems can be designed with controllable, non-reciprocal chemotactic interactions to generate emergent self-organization and collective behaviors reminiscent of biological systems.

**One sentence summary:** Chemotactic droplets chase and communicate by nonreciprocal oil exchange resulting in emergent self-organization.

**Main text:**

The emergent behaviors, dynamics, and patterns that evolve within multibody systems such as swarming insects, crowds of people, or bacterial colonies result from non-equilibrium, distance-dependent coupling between group constituents that is often nonreciprocal, involving both attractive and repulsive interactions(*1-3*). An example two-body nonreciprocal interaction is that of predator and prey, where the predator is attracted to the prey, but the prey is repelled by the predator. In living systems, complex chemomechanical feedback networks govern these nonreciprocal interactions and mediate organisms' abilities to sense, respond, and interact with neighbors. However, similar requisite control over the strength and asymmetry of such nonreciprocal interactions in inanimate systems(*4-6*) has not been possible to achieve, likely because chemical gradients not only serve to signal communication between group constituents but also provide the driving force for motion (e.g. chemotaxis)(*7*). Here, we report an experimental framework by which to chemically program the specificity, directionality, and strength of nonreciprocal predator-prey-like chasing and multibody interactions between microscale oil-in-water droplets. The interactions and dynamic self-organization result from the net directional, micelle-mediated transport of oil between emulsion droplets of differing composition and are powered by the free energy of mixing. The nonreciprocal behavior occurs in a wide variety of oil and surfactant conditions, and we systematically elucidate chemical design rules for tuning the interactions between droplets by varying oil and surfactant chemical structure and concentration. Experimentally determined two-body interaction parameters successfully predict and reproduce multibody behavior in simulations, suggesting that emulsions can serve as a tractable experimental framework for the study of active materials capable of adaptive, life-like, non-equilibrium dynamics. This robust but facile platform lays the groundwork for design of soft materials with motility-induced self-organization governed by chemical transport and provides a simple physical model for studying emergent, collective phenomena in systems with minimal molecular complexity.

Emulsions, which are phase-separated dispersions of fluids stabilized by surfactants, have recently emerged as a rich materials platform in which to study non-equilibrium chemotactic interactions(*8*). Emulsions persist in a thermodynamically out-of-equilibrium state and are highly dynamic, with molecules being continually exchanged between droplets and the continuous phase(*9*). Droplets can also chemotax, moving in response to chemically-induced interfacial tension gradients through Marangoni flow(*10-14*). Building upon the recent discovery that micelle-mediated oil solubilization gradients generate reciprocal repulsion between like droplets(*15*), we wondered whether asymmetric interactions would occur between droplets of oils with very different micellar solubilization profiles. Thus, we first investigated the interactions between microscale droplets of 1-bromooctane (BOct) and ethoxynonafluorobutane (EFB), two fully miscible, dense oils with very low water solubility and anticipated differences in oil-micelle interactions due to varying degrees of fluorination(*16*) (**Figure 1A**). In 0.5 wt% aqueous Triton X-100 nonionic surfactant (Triton), a concentration over 25 times Triton's critical micelle concentration (CMC), we observed that not only were the droplet interactions asymmetric, but they were also nonreciprocal (**Figure 1A, Video S1**). Even at initial droplet separation distances of tens of microns, BOct "predator" droplets were attracted to, and accelerated towards, EFB "prey" droplets, and the EFB prey were repelled by the approaching BOct predators (**Figure 1A**). These chases led to the formation of droplet pairs or many-droplet clusters propelled at speeds of upwards of 20 μm/s with sustained interactions lasting over an hour (**Figure 1B, Video S2**). No chasing occurred in control experiments in which either the EFB or BOct droplets were replaced with similarly sized polystyrene particles, and isolated droplets of BOct and EFB also did not display active behavior (i.e. individual droplets were not self-propelled). Because BOct droplets solubilized into the continuous phase at a significantly faster rate than did EFB droplets (**Figure 1C**), these nonreciprocal chasing interactions appeared to be primarily driven by gradients of BOct. Droplet motion was sustained as long as BOct predator droplets were not completely solubilized, which usually took over an hour. In the presence of EFB drops, BOct was not only solubilized into the continuous phase, but it was also transferred into the EFB drops as evidenced by an increase in EFB droplets' refractive index over time (**Figure 1D**). Pre-saturating the surfactant solution with both oils prior to adding both droplets still resulted in chasing interactions. Hence, it appears that the nonreciprocal chasing behavior is related to asymmetric oil transport between droplets, where BOct undergoes micellar solubilization into the continuous phase and is subsequently transferred from the micelles into EFB droplets, while transfer of EFB in the reverse direction is minute (**Figure 1E**).

In order to test this directional transport hypothesis, we aimed to quantify the energies associated with chasing droplet pairs and examine how these energies vary as a function of parameters that influence micellar transport, such as surfactant concentration and chemical structure. At low Reynolds numbers, for any interaction between two droplets with separation distance $r$, the area under the curve $\delta r/\delta t$ versus $r$ (shaded grey region, **Figure 2A**) is proportional to the interaction energy. A negative interaction energy indicates a net attraction (e.g. a successful chase, where the predator catches the prey), while a positive interaction energy indicates repulsion (e.g. a failed chase, where the predator does not catch the prey). Because the droplet interactions are nonreciprocal, not only does $r$ change over time, but the midpoint between the droplets also moves with velocity $V_{midpoint}$ as the predator approaches the prey. We use the area under the $V_{midpoint}$ versus $r$ curve (shaded purple region, **Figure 2A**), which is proportional to the displacement energy, to quantify the degree of nonreciprocity. As defined in **Figure 2A,** a positive displacement energy indicates BOct chases EFB, while a negative displacement energy corresponds to the reverse chasing direction. By multiplying the areas under these two curves in **Figure 2A** by the Stokes' drag and introducing a dimensionless constant $C$ to account for contributions to the drag constant from droplet-internal flows and substrate proximity effects, we could calculate the interaction and displacement energies for any two-body droplet interaction. For chasing BOct and EFB droplets in 0.5 wt% Triton, these energies are on the order of $10^{-18}$ J, or $10^4$ $k_BT$ where $k_B$ is the Boltzmann constant and $T$ is temperature in Kelvin (taken as 298 K) (**Figure 2B-D**). Please refer to SI section "Quantification of interaction energy and displacement energy for two-body droplet chasing" for further details of these calculations.

Having established a method by which to quantitatively compare the interaction and displacement energies associated with any two-body chasing interaction, we examined how those energies change with surfactant concentration, surfactant chemical structure, and oil chemical structure. At each set of experimental conditions, we analyzed multiple (at least 3) independent chasing interactions. As concentrations of Triton increased from 0.1 wt% to 0.5 wt% (which are all concentrations above the CMC), the magnitude of both the interaction and displacement energies associated with EFB and BOct encounters increased, indicating that a higher concentration of micelles is associated with a greater flux of BOct and stronger chasing interactions (**Figure 2B**). Other surfactants that preferentially solubilize BOct, including both ionic (e.g. sodium dodecyl sulfate at 3 wt%) and nonionic surfactants (e.g. poly(ethylene glycol) (12) tridecyl ether at 1 wt%) generated similar chasing behavior. Replacing Triton with 3.0 wt% Capstone FS-30 (Capstone), a nonionic fluorosurfactant that preferentially solubilizes the fluorinated EFB, reversed the chasing direction such that EFB was the predator and BOct the prey (**Figure 2C, Video S3**). This reversal in chasing direction is indicated by the sign reversal of the displacement energy (**Figure 2C**) and shows that the net direction of micelle-mediated oil transport determines the chasing direction. Mixing 0.5 wt% Triton and 3.0 wt% Capstone in varying ratios allowed continuous tuning of the chasing speed and direction of BOct and EFB pairs (**Figure 2C**). If both oils solubilized at appreciable rates in a surfactant mixture (e.g. **Figure 2C(ii)**) then no chasing occurred, which is consistent with the hypothesis that net directional oil transport between droplets is required. When EFB was replaced with methoxyperfluorobutane, another fluorinated oil having partial miscibility with BOct, similar chasing interactions in 0.5 wt% Triton occurred; however, when EFB was replaced with perfluorohexane, a fluorinated oil with very little miscibility with BOct, there was no chasing (**Figure 2D**). Because none of the fluorinated oils solubilize appreciably in 0.5 wt% Triton, this influence of oil miscibility on chasing suggests that favorable oil mixing is also essential for the predator-prey interaction to occur.

Our observations regarding the influence of micelle-mediated oil transport on the chasing interactions suggest the following underlying Marangoni-flow-driven mechanism for nonreciprocal droplet behavior (**Figure 2E):** 1) Predator drops produce oil-filled micelles that repel *all* droplet neighbors because the oil solubilizate decreases the surfactant's ability to stabilize oil-water interfaces (**Figure S1**). Gradients of oil-filled micelles thus cause an asymmetric surfactant distribution on the surface of neighboring droplets, driving motion away from the predator's signal via Marangoni flow. 2) Prey droplets uptake the predator's solubilized oil from micelles, "freeing" surfactant molecules that can more efficiently stabilize oil-water interfaces. The oil uptake by the prey thus attenuates the solubilizate gradients and attracts predators. This experimental description is analogous to the "source" and "sink" framework often used to achieve predator-prey interactions in theoretical models(*17*); here, the BOct drop is the "source", (e.g. an emitter of the chemical signal), and the EFB drop is the "sink" (e.g. a consumer of the chemical signal). The "source" and "sink" act together to modify the chemical gradients of solubilizate that modulate interfacial tension gradients, creating a mechanism for chemo-mechanical feedback leading to nonreciprocal behavior.

To test the generality of this proposed mechanism amongst oils of more similar chemistry, we systematically examined the chasing direction and speed of pairs of 1-iodo-*n*-alkane micro-droplets (**Figure 3A**). Iodo-*n*-alkanes (hereafter identified by carbon number, $n = 4$ to 16) were chosen instead of *n*-alkanes or 1-bromo-*n*-alkanes for ease of experimentation because they are denser than the aqueous continuous phase and remain liquid at room temperature for all chain lengths up through 1-iodohexadecane. Chasing between droplets of many oil combinations was observed (**Figure 3A**), and interestingly, even a structural difference as minute as a single methylene bridge (-$CH_2$-) could be sufficient to drive chasing (e.g. 1-iodoheptane chases 1-iodohexane, **Video S4** and **Figure S2**). By dispersing tracer particles in each of the oils, we directly visualized the distinct Marangoni flow patterns formed in both predator and prey droplets as they chased (**Video S5**). As highlighted by the green and yellow regions in **Figure 3A**, the designation of predator and prey for each pairing was not simply a function of relative oil chain length; oil droplets of both the shortest ($n = 4$ to 6) and longest ($n = 12$ and 16) chain lengths tended to be prey while intermediate chain lengths ($n = 7$ to 10) tended to be predators (**Figure 3A**). Hydrocarbon oils exhibited qualitatively similar chasing trends as a function of chain length (e.g. octane chased both hexane and hexadecane in 0.5 wt% Triton, **Figure S2C**). Oil solubilization rates for isolated droplets (**Figure 3B**) also did not decrease monotonically with increasing *n* as would be expected in the absence of surfactant(*18*), suggesting that oil solubilization kinetics, and hence the chasing direction, are influenced by two transport pathways(*9, 19, 20*): 1) direct water dissolution (dominant for oils with lower *n* and higher water solubility) and 2) direct micellar transfer (dominant for oils with higher *n* and lower water solubility). Droplets of oils for which transfer is dominated by pathway #1 were always prey (e.g. $n = 4$ and 5) even if they solubilized faster than their predators. If solubilization of both oils in a pair was dominated by the micellar transfer of pathway #2 (i.e. $n = 7$ and higher), then the relative rates of oil solubilization did qualitatively correlate with chasing direction and the predator was always solubilized more rapidly than the prey. If neither oil in a pair solubilized appreciably by pathway #2, then the interactions were weak to nonexistent. These observations lead to an important insight: the direction and speed of chasing is influenced not only by the direction and rate of net oil transport, but also by the specific molecular pathway for that transport, where pathway #1 does not contribute as significantly to the Marangoni effect driving chasing as does pathway #2. Accordingly, the highest chasing efficiencies (**Figure 3C**, see SI section "Efficiency of chasing driven by oil exchange" for a description of the calculation) were found for oil pairs where the predator droplets were preferentially solubilized by the direct micellar transport pathway #2 (e.g. $n = 7$ or higher) and where the prey did not solubilize appreciably ($n = 16$) so as to maximize the asymmetry of the interaction and minimize total oil lost to the continuous phase. Maximum efficiencies obtained for the iodoalkane droplet pairs were on the order of $10^{-7}$, which is approximately two orders of magnitude higher than the efficiency of well-studied, self-electrophoretic bimetallic motors(*21*).

In experiments with a high number density of droplets, the formation of chasing pairs was the first step in a cascade of collisions and reorganizations resulting in the assembly of myriad clusters with diverse structures and dynamics (**Figure 4A, Video S6**). Multibody interactions between BOct and EFB oil droplets in 0.5 wt% Triton generated clusters that rotated, translated or remained stationary depending on their geometry (**Figure 4A**). A short-ranged sticking attraction between droplets(*22, 23*), led to the formation of clusters that would otherwise have been unstable due to multiple prey droplets fleeing in different directions from the same predator (e.g. the clusters bordered by dashed lines in **Figure 4A**). While varying the droplet sizes did not qualitatively change the two-body chasing direction, droplet size did affect the multibody behavior. Not only did the relative diameter of predator and prey impact packing geometry of the clusters, but BOct droplets with diameters below approximately 40 μm diameter became self-propelled(*12*) (**Figure S3**) and would continue to translate around the EFB drop to which they were attached. Such predator self-propulsion resulted in trochoidal trajectories (for two-body interactions, SI section "Trochoidal trajectories"), and in flapping and run-and-tumble dynamics (for multibody clusters) (**Figure 4B, Video S7**). To investigate the extent to which the multibody cluster dynamics can be understood by summation of the measured two-body droplet interactions, we compared our experiments with simulations (see SI section "Quantification of interaction parameters for simulations" for details). The circular and linear motions of clusters could be readily reproduced in simulations by using the initial positions of the droplets and experimentally-determined two-body interaction parameters as inputs (**Figures 4C,D** and **Videos S8-S10**). Due to long-range solute-mediated interactions, neighboring clusters interact and influence each other's structure and trajectory. For example, both experiment and simulation show that individual droplets can act as chemical signaling posts to effectuate turns and reorientations of nearby chasing pairs (**Figure 4D, Video S9**) or cause neighboring clusters to disassemble (**Video S10**). Thus, in contrast to static, self-assembled equilibrium structures, these droplet clusters were dynamically disassembled, rearranged, and reformed over time through non-equilibrium self-organization mediated by solute gradients and physical collisions(*24*) (**Figure 4E, Video S6**).

We have established a facile, robust platform for producing controllable, nonreciprocal predator-prey interactions between oil-in-water droplets powered by the free energy of mixing and achieved by combining micellar oil exchange and Marangoni flow. Only four chemical components are necessary to tune the predator-prey interactions in this system: two chemically distinct but miscible oils, a surfactant that preferentially solubilizes one of the oils in micelles, and water. The chasing speed, direction, and efficiency are governed by the rate and direction of micelle-mediated oil exchange between droplets, which can be rationally controlled by varying the oil chemical structure, surfactant chemical structure, or surfactant concentration. Many-body interactions were accurately described by simulations based on measured two-body droplet interactions, suggesting that this experimental platform may be suitable for testing theoretically predicted aspects of dynamic collective behavior. Further routes to control droplet chasing through tailoring of solubilizate-micelle intermolecular interactions(*25*) may enable selective chemical transport across multiple orthogonal communication pathways for more complex networks of interacting droplets. We expect that this experimental platform and the concomitant chemical design rules laid forth will pave the way to developing inanimate systems with dynamic or life-like adaptive organizations and functionalities(*26, 27*) that are not accessible through equilibrium assembly.

**Acknowledgments.** The authors thank Ciera Wentworth for her help obtaining interfacial tension measurements using the pendant drop method. L.D.Z., C.H.M. and Y-J.C. acknowledge support from the Army Research Office through grant number W911NF-18-1-0414 and the Penn State MRSEC funded by the National Science Foundation (DMR-1420620). C.H.M. acknowledges support from the Thomas and June Beaver Fellowship at Penn State and the Pennsylvania Space Grant Fellowship and Y-J.C. received support from the Erickson Discovery Grant program at Penn State. P.G.M. acknowledges funding from the NWO (Dutch national science foundation) Graduate Program through the Debye Institute for Nanomaterials. J.G. acknowledges support from the Guangdong Innovative Research Team Program (No. 2011D039).

**Author contributions.** C.H.M., P.G.M., and L.D.Z. developed the concept for the research. C.H.M and Y-J.C. conducted experiments of droplet chasing under different conditions and measured oil solubilization rates. P.G.M. analyzed data on the droplet motions and performed the simulations. C.H.M., P.G.M., and L.D.Z. wrote the manuscript. P.G.M. was supervised by W.K.K., J.G. and A.v.B. C.H.M. and Y-J.C. were supervised by L.D.Z. All authors discussed the results and manuscript.  Authors declare no competing interest. "All data is available in the main text or the supplementary materials."**Supplementary Materials:**
Materials and Methods
Figures S1-S6
Table S1
Movies S1-S10
References (28-30)

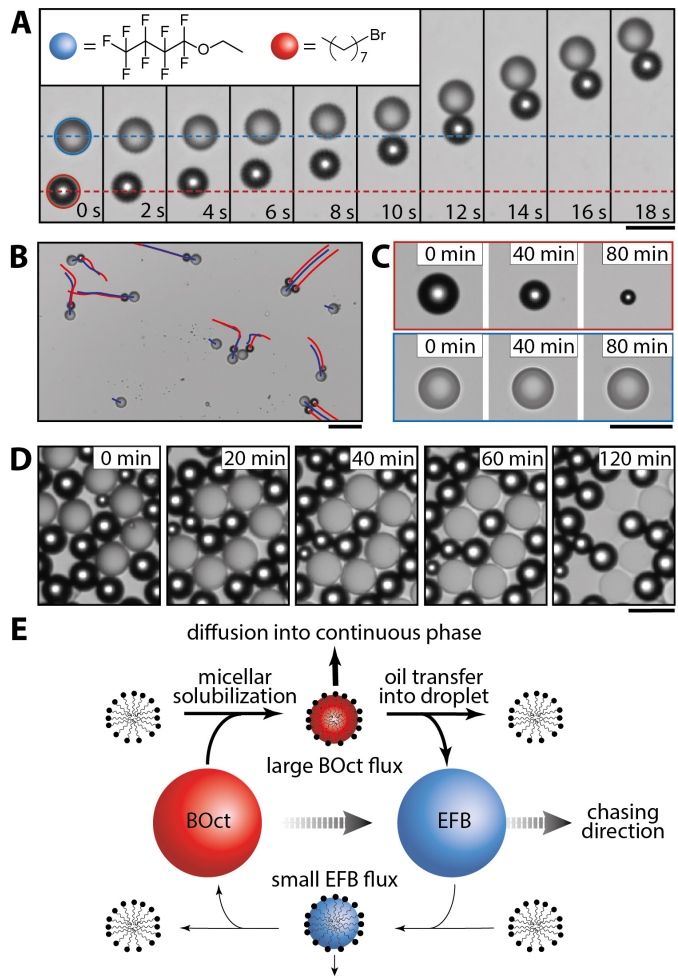

**Figure 1**. **Bromooctane (BOct) droplets chase ethoxynonafluorobutane (EFB) droplets in aqueous surfactant due to micelle-mediated oil transport.** **A**, Time-sequence frames of a chasing interaction between a BOct predator droplet (outlined in red) and an EFB prey droplet (outlined in blue) in 0.5 wt% Triton X-100 (Triton) aqueous solution (as seen in **Video S1**). BOct accelerates towards the EFB, and the EFB moves away in response. Eventually, the EFB is "caught" as the two droplets touch and continue to translate as a pair. Scale, 100 μm. The droplets' initial positions are indicated by the colored, dashed lines. **B**, Trajectories of BOct drops (red lines) and EFB drops (blue lines) over a period of thirty seconds shows the linear motion of pairs and multi-droplet clusters while individual droplets are stagnant. Scale, 250 μm. **C**, Optical micrographs of an isolated BOct droplet (top) and EFB droplet (bottom) in 0.5 wt% Triton over time. The diameter of BOct droplets decreased at a rate of 0.49 μm/min due to micellar solubilization, but the EFB droplets did not solubilize appreciably. Scale, 100 μm. **D**, When EFB and BOct droplets were mixed together in 0.5 wt% Triton, the refractive index of EFB ($n$ = 1.28) increased over time from uptake of the mobile, higher refractive index BOct ($n$ = 1.45), eventually becoming nearly transparent due to index matching with the aqueous phase ($n$ = 1.33) and swelling slightly in size. Scale, 100 μm. **E**, We hypothesize that the nonreciprocal chasing interactions are associated with the asymmetric rates of micelle-mediated oil exchange between droplets, where the flux of BOct is much greater than that of the EFB in 0.5 wt% Triton due to the BOct's higher micellar solubilization rate.

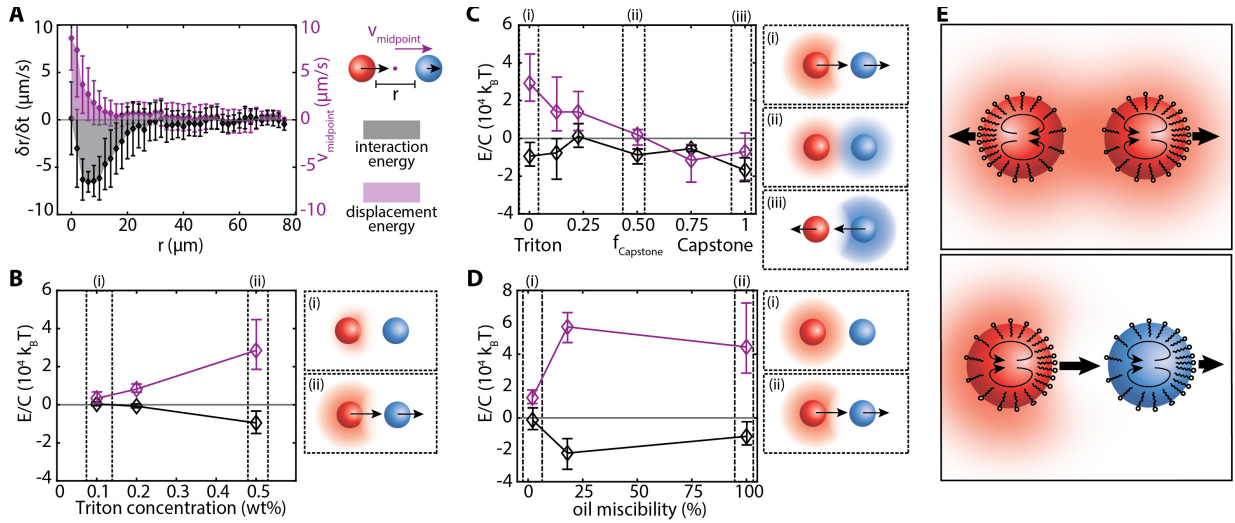

**Figure 2. The interaction and displacement energies associated with predator-prey droplets are tunable by varying the chemical structure and concentration of surfactant and oil miscibility. A,** Interaction energy and displacement energy can be calculated for any two-body interaction from measurements of the distance between droplets, $r$, and the velocity of the midpoint between droplets, $V_{midpoint}$, as defined in the schematic. Specific data shown are for the interaction between BOct (red drop) and EFB (blue drop) in 0.5 wt% Triton, and each data point represents the average and standard deviation from at least 3 independent droplet encounters obtained from multiple experiments under the same conditions. The interaction energy between droplets for any given two-body interaction is proportional to the integrated area under the curve $\delta r/\delta t$ vs. $r$ (grey shaded region). A negative interaction energy indicates net attractive interaction (e.g. a successful chase), while a positive interaction energy indicates net repulsion (e.g. a failed chase). The displacement energy is proportional to the area under the $V_{midpoint}$ vs. $r$ curve (shaded purple region). As defined, a positive displacement energy indicates that BOct chases EFB, while a negative displacement energy corresponds to the opposite chasing direction. **B,** The interaction energy (black line) and displacement energy (purple line) were measured for droplet chasing under varying concentrations of Triton by creating plots like that shown in (**A**) for each droplet encounter; the shaded grey and shaded purple areas were multiplied by a drag constant, which we take to be the Stokes' drag multiplied by a dimensionless factor ($C$) to account for contributions to the drag constant from droplet-internal flows and substrate proximity effects, yielding energies on the order of $10^4$ $k_BT$ where $k_B$ refers to the Boltzmann constant and $T$ refers to the temperature in Kelvin, which we take to be 298 K. Please refer to SI section "Quantification of interaction energy and displacement energy for two-body droplet chasing" for calculation details. Each data point represents an average from three droplet encounters with error bars representing highest and lowest measurements. Schematics at right illustrate the production of solubilizate gradients around each drop and the resultant chasing directions for regions (i and ii) where BOct is the red drop and EFB is the blue drop. At 0.1 wt% Triton, the solubilization rate of BOct was low (0.09 μm/min change in diameter) and droplet interactions were weak (region (i)). At higher concentrations of Triton, the solubilization rate of BOct increased (0.49 μm/min change in diameter, region (iii)) and interactions were stronger. **C,** Interaction and displacement energies for BOct and EFB as a

function of $f_{Capstone}$, which corresponds to the volume fraction of 3 wt% Capstone FS-30 (Capstone) in a mixture with 0.5 wt% Triton. Capstone is a nonionic fluorosurfactant that at 3.0 wt% solubilizes EFB at a rate of diameter change of 0.34 μm/min and BOct rate of 0.08 μm/min. Schematics at right represent the relative oil gradients around each drop and chasing directions for regions (i-iii) where BOct is the red drop and EFB is the blue drop. Notably, the chasing direction reverses as the volume fraction of Capstone increases as evidenced by the displacement energy's sign change. **D**, Displacement and interaction energies as a function of prey fluorinated oil miscibility with predator BOct in 0.5 wt% Triton. The three fluorinated oils tested were perfluorohexane (0.5% miscibility with BOct, region (i)), methoxyperfluorobutane (17% miscibility with BOct), and EFB (100% miscibility with BOct, region (ii)). Please refer to Methods Section, "Measurement of oil miscibility" for description of miscibility determination. Schematics at right illustrate the relative oil gradients and chasing directions for regions (i and ii) where BOct is the red drop and the fluorinated oil is the blue drop. Chasing occurred for BOct paired with either methoxyperfluorobutane or EFB due to sufficiently high miscibility, but no chasing occurred with perfluorohexane, which had insufficient miscibility. **E,** Schematic illustration of how micelle-mediated oil transport couples to Marangoni flow driving droplet motion. Top and bottom: predator oil droplets (red) rapidly produce oil-filled micelles and repel all droplet neighbors because the oil solubilizate decreases the surfactant's ability to stabilize oil-water interfaces. Bottom: prey oil droplets (blue) uptake the predator's solubilized oil from micelles, attenuating the oil gradients and attracting predators. The prey is still repelled by the predator, however, leading to the nonreciprocal chasing interaction.

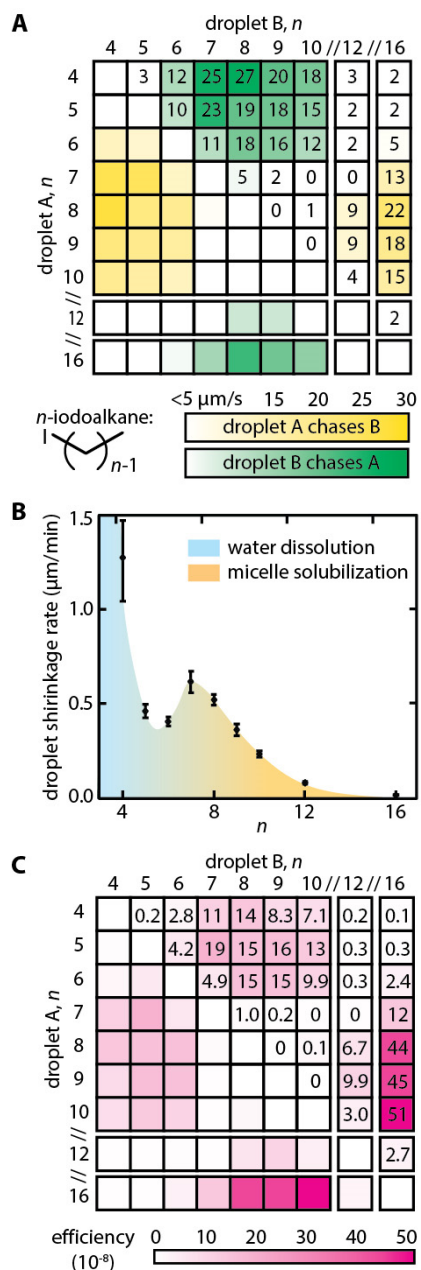

**Figure 3. Oil molecular structure influences the direction, speed, and efficiency of chasing between droplets. A,** Tabulated average chasing speeds in 0.5 wt% Triton for all pairwise combinations of 1-iodo-$n$-alkane droplets, identified by carbon number $n$ = 4 to 16. Chase speed is defined as the speed of the predator-prey pair once contact between droplets was made, and the speed reported is the average of three chases. Color coding in the yellow and green regions indicates the chasing direction between droplet A and droplet B of defined carbon number $n$. Pairs for which interactions were weak (with speeds below 5 µm/s) or reciprocal (left to right center diagonal) were left white. **B,** Solubilization rates (change in drop diameter over time) of the iodo-$n$-alkanes were measured for isolated droplets in 0.5 wt% Triton. For each value of $n$, the solubilization rate of three independent droplets was measured and the average, maximum, and minimum are reported in the plot. Two solubilization pathways exist: direct water dissolution (cyan region, more likely for oils with lower $n$ and higher water solubility) and

micellar transfer (orange region, more likely for oils with higher $n$ and lower water solubility). **C,** The efficiency of droplet chasing for each pair of iodo-*n*-alkanes was calculated based on the combined rates of oil loss for predator and prey (i.e. the chemical energy input) and the resulting speed of the chasing droplet pair (i.e. the mechanical energy output) (see SI section "Efficiency of chasing driven by oil exchange" for calculation details). Highest efficiencies (dark pink areas) were found for oil pairs in which the predators primarily solubilized by the micellar transfer pathway and the prey did not solubilize appreciably as to minimize oil lost to the water and maximize the asymmetry of the predator-prey interaction.

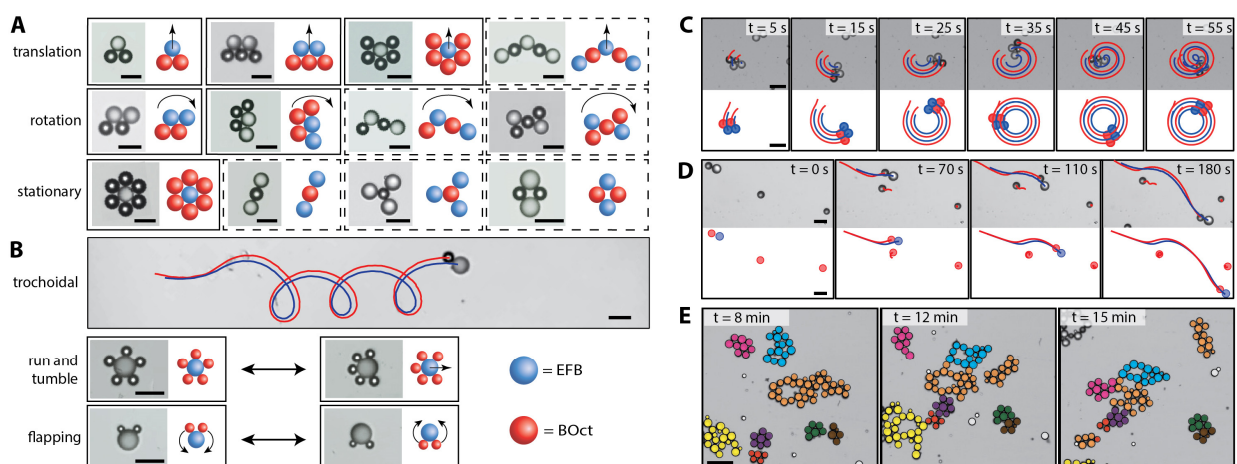

**Figure 4. Multibody nonreciprocal interactions between droplets cause emergent assembly and disassembly dynamics that are predictable based on measured two-body interactions.**
**A**, In mixtures containing a high number density of BOct and EFB droplets in 0.5 wt% Triton, multi-droplet clusters assembled via nonreciprocal interactions and exhibited movements (indicated by arrows) dependent on cluster geometry. A short-ranged sticking attraction between droplets led to the formation of cluster geometries that would be otherwise unstable to formation (bordered by dashed lines). Scale, 100 μm. **B**, As the diameter of BOct decreased below approximately 40 μm, the droplets became self-propelled and started to migrate along the surface of the prey, leading to the formation of trochoidal trajectories, run-and-tumble dynamics, and flapping motions. Scale, 100 μm. See SI section "Trochoidal trajectories" and **Video S7**. **C,** Simulations reproduce the observed cluster dynamics of EFB (blue) and BOct (red) using the initial droplet positions and interaction parameters obtained from their two-body interaction potential measurements (see section "Quantification of interaction parameters for simulations" in SI for specific details). Shown is the rotation of a four-droplet cluster containing two predators and two prey (**Video S8**). Scale, 100 μm. **D,** Solute-mediated interactions operate at distances of upwards of 50 μm, causing individual droplets (here, methoxyperfluorobutane) to act as signaling posts that direct the movement of a nearby chasing pair of BOct and methoxyperfluorobutane as shown in experiment and simulation (**Video S9**). Scale, 100 μm. **E,** Droplets of BOct and EFB continuously assemble and reorganize over time in 0.5 wt% Triton (**Video S6**). Droplets in the same cluster at $t$ = 8 minutes are labeled in the same color, isolated droplets not in a cluster at $t$= 8 minutes are labeled in white, and droplets not in the frame at $t$ = 8 minutes but which entered the field of view at later time points were unlabeled. As evidenced by the color mixing, these clusters were continuously rearranged and reformed from collisions and solution-mediated interactions. Scale, 200 μm.

# Supplementary Materials for

# Predator-Prey Interactions between Droplets Driven by Nonreciprocal Oil Exchange

Caleb H. Meredith[1†], Pepijn G. Moerman[2,3†], Jan Groenewold[2,4], Yu-Jen Chiu[1], Willem K. Kegel[2], Alfons van Blaaderen[3], Lauren D. Zarzar[1,5,6]*

1. Department of Materials Science and Engineering, The Pennsylvania State University, PA 16802, USA
2. Van't Hoff Laboratory for Physical & Colloid Chemistry, Debye Institute for Nanomaterials Science, Utrecht University, The Netherlands
3. Soft Condensed Matter, Debye Institute for Nanomaterials Science, Utrecht University, The Netherlands
4. Guangdong Provincial Key Laboratory of Optical Information Materials and Technology & Institute of Electronic Paper Displays, South China Academy of Advanced Optoelectronics, South China Normal University, Guangzhou 510006, P. R. China.
5. Department of Chemistry, The Pennsylvania State University, PA 16802, USA
6. Materials Research Institute, The Pennsylvania State University, PA 16802, USA

† These authors contributed equally to this work

Correspondence to: ldz4@psu.edu

**This PDF file includes:**

> Materials and Methods
> Supplementary Text
> Figs. S1 to S6
> Table S1
> Captions for Movies S1 to S10

**Other Supplementary Materials for this manuscript include the following:**

> Movies S1 to S10

**Materials and Methods**

Chemicals. All chemicals were used as received. Capstone FS-30, 1-(ethoxy) nonafluorobutanes (mixture of n- and iso-butyl isomers), perfluorohexane(s) (98%), (Synquest Laboratories); methoxyperfluorobutane (mixture of n- and iso-butyl isomers) (99%), Triton X-100, poly(ethylene glycol) (12) tridecyl ether, sodium dodecyl sulfate (99%), 1-iodoheptane (copper stabilized, 98%), 1-iodononane (copper stabilized, 98%), *n*-hexane (>99%) (Sigma Aldrich), 1-bromooctane (98%), 1-iodobutane (copper stabilized, 99%), 1-iodopentane (copper stabilized, 98%), 1-iodohexane (copper stabilized, 98%), 1-iodooctane (copper stabilized, 98%), 1-iododecane (copper stabilized, 98%), 1-iodododecane (copper stabilized, 98%), 1-iodohexadecane (copper stabilized, 98%), *n*-octane (98%), (Alfa Aesar), hexadecane (>98%), (TCI), Lumogen F Red 305 (BASF), hydrophobic iron oxide nanoparticles (CMS Magnetics ferrofluid, particle size unspecified, purchased on Amazon).

Droplet fabrication. Fabrication of monodisperse emulsion droplets was carried out using a four-channel flow focusing glass hydrophilic microfluidic chip with a 100 μm channel depth (Dolomite). Each microchannel inlet was connected to a reservoir of the desired liquid. The inlets for the inner fluid phases (for example bromooctane or ethoxynonafluorobutane) were connected to the reservoirs with inner diameter 0.0025 inch (63.5 μm), outer diameter 1/16 inch (1.59 mm) polyether ether ketone (PEEK) tubing 20 inches (50.8 cm) in length, and the outer fluid phase containing aqueous surfactant solution (for example 0.5 wt% Triton) was connected to the reservoirs with inner diameter 0.005 inch (127 μm), outer diameter 1/16 inch (1.59 mm) PEEK tubing 20 inches (50.8 cm) in length. The flow rates of each liquid were manipulated with a Fluigent MFSC-EZ pressure controller in order to control droplet size. Typical pressures used for the inner phase fluids ranged from 1.0 bar to 3.0 bar and typical pressures for the outer phase fluids ranged from 0.50 bar to 6.0 bar.

Droplet mixing protocol for observing two-body and multibody droplet interactions. To observe and analyze droplet chasing interactions, a glass bottomed dish of 1.5 inch (3.8 cm) diameter with aluminum walls was filled with approximately 2 mL of surfactant solution and then approximately 100 prefabricated oil droplets of each type were transferred from a storage vial via micropipette. The number of droplets was chosen to produce a dilute layer of droplets along the dish bottom (area fraction < 0.001) necessary for analyzing isolated chasing interactions between pairs of droplets. (The exception is the experiment shown in **Video S6**, where we examined the interactions among a high density of droplets). The solution was then gently agitated for several seconds to randomly disperse the oil droplets throughout the dish. After dispersing the droplets, the dish was covered with a lid and left undisturbed for the remainder of the experiment. Droplets sedimented to the dish bottom within several seconds. Chasing interactions began within one minute and continued to grow in number over the course of the experiment. The formation of larger multi-droplet clusters occurred more gradually over a period of minutes but continued to grow in size as long as droplets remained active. Droplet motion and chasing interactions persisted as long as predator droplets were not completely solubilized (typically on the order of an hour). By choosing oils that solubilized more slowly, or by starting with larger droplets, motion could be extended for at least several hours before predator droplets were solubilized completely. Droplets in stored in saturated surfactant solutions for more than a week still exhibited chasing interactions when added to fresh surfactant solutions. Chasing interactions also still occurred even if the surfactant solution was pre-saturated with the two oils.

Droplet imaging. Droplet interactions were observed using an inverted optical microscope (Nikon, Eclipse Ti-U) in brightfield transmission mode between 2x and 40x magnification. Images were recorded using an attached camera (Andor, Zyla sCMOS) in video mode ranging from 0.1 to 25 frames per second. To distinguish droplets of similar refractive index and verify chasing direction, Lumogen F Red 305 fluorescent dye was added to one droplet type and an overlay of brightfield and fluorescence imaging was used (e.g. **Figure S2**). Controls were conducted to verify that the dye did not affect the chasing dynamics.

Observation of oil exchange. To observe the exchange of oil between BOct and EFB droplets as shown in **Figure 1D**, a single depression microscope slide (AmScope) was completely submerged in a large dish containing 0.5 wt% Triton X-100 surfactant solution. An approximately equal number of BOct and EFB droplets of about 70 µm diameter were added via micropipette to the center of the slide depression where they were spatially confined in a circular monolayer of 1 cm diameter. Droplets were then monitored for a period of more than four hours. Over time, the EFB droplets (refractive index, $n$ = 1.282) visually disappeared due to index matching with the surrounding aqueous phase ($n \approx 1.33$) caused by an uptake of the higher index BOct oil ($n$ = 1.453). Over a similar time period, BOct droplets across the cluster were found to shrink noticeably as their oil was more readily solubilized by the Triton surfactant micelles.

Measurement refractive indices and oil miscibility. The miscibility of fluorinated oils and BOct was analyzed based on refractive index measurements performed using a temperature-controlled J457FC refractometer (Rudolph Research) at 20 °C. BOct (refractive index $n$ = 1.453, measured; $n$ = 1.450 to 1.463, lit.(28) at 20 °C) and EFB ($n$ = 1.282, measured) were completely miscible at room temperature, and the refractive index of their binary mixtures varied linearly as a function of volume fraction of each oil. Given this observation, we presumed that a similar linear trend would hold for binary mixtures of other hydrocarbon and fluorocarbon oils and that refractive index could be used to quantify degree of miscibility by extrapolating between the indices of the two pure oils. Methoxyperfluorobutane ($n$ = 1.271, measured) was saturated with BOct and the refractive index of the mixture was measured ($n$ = 1.302), yielding an estimated miscibility of $\frac{1.302-1.271}{1.453-1.271} = 0.17$, or 17%. Perfluorohexane ($n$ = 1.254, measured; $n$ = 1.252, lit.(28) at 22°C) was saturated with BOct and the refractive index of the mixture was measured ($n$ = 1.255), yielding an estimated miscibility of $\frac{1.255-1.254}{1.453-1.254} = 0.005$, or 0.5%.

Observation of Marangoni Flow. The Marangoni-driven flows inside chasing iodoheptane and iodohexane droplets were visualized by dispersing 0.5 vol% hydrophobic iron oxide nanoparticle solution (see Materials) in the oil phase prior to emulsification. This combination of oils was chosen because they were found to readily disperse the hydrophobic iron oxide nanoparticles. The particles were dispersed in each oil prior to the emulsification of droplets using microfluidics. After emulsification, the droplets were mixed in 0.5 wt% Triton X-100 surfactant solution following the protocol described above in the section, "Droplet mixing protocol for observing two-body and multibody droplet interactions". The iodoheptane droplets were found to chase the iodohexane droplets, similar to the experiments without iron oxide particles (**Figure S2)**, suggesting that the iron oxide particles did not have a significant affect on the droplet interfaces and were inert tracers of flow. When the two types of droplets chased each other, asymmetric flow profiles were observed in both droplets (**Video S5**), consistent with Marangoni flow as diagrammed in **Figure 2E**.

Oil solubilization rate measurement. Oil solubilization rates were determined by monitoring the sizes of individual isolated droplets (starting diameter, 60 to 80 μm) in different surfactant solutions over a period of at least 20 minutes. Time lapse images were analyzed using calibrated pixel values using Nikon Elements-D software suite. The rate of droplet diameter change was found to be constant over the observation period indicating an interfacially limited solubilization process which is consistent with previous kinetic solubilization studies on nonionic surfactants(*19*). Solubilizing droplets smaller than 40 μm often became self-propelled as characterized by their persistent directional motion even in the absence of other surrounding droplets (**Figure S3**). Solubilizing droplets have been previously studied and exhibit self-propelled motion spontaneously above a critical Péclet number(*11*, *12*). Due to their advection, self-propelled droplets were found to solubilize faster than stationary droplets and were therefore excluded from measurements of solubilization rate.

Interfacial tension measurement. Oil-water interfacial tensions were obtained by the pendant drop method using a Ramé Hart 250 Automatic Goniometer. A 28-gauge stainless steel probe tip needle was used to dispense oil drops into a glass cuvette filled with approximately 5 mL of aqueous surfactant solution. DROPimage Advanced analysis software package was used to analyze images of drop shape to determine interfacial tension values based on the measuring the curvature deformation of the drop in response to gravitational and buoyant forces. Interfacial tension of BOct and EFB was measured in 0.5 wt% Triton solution, as well as in BOct-saturated 0.5 wt% Triton solution. The BOct-saturated surfactant solution was prepared by emulsifying an excess volume of BOct oil (>5 vol%) in 0.5 wt% Triton and allowing it to equilibrate for more than 72 hours before measurement. The BOct oil for all experiments was itself was also pre-equilibrated with 0.5 wt% Triton for one week prior to interfacial tension measurement to account for possible surfactant partitioning into the oil(*29*). A single dynamic interfacial tension measurement curve is shown for each condition (**Figure S1**). At least three measurements were conducted for each condition, and the observation of elevated interfacial tension values for the BOct-saturated solutions was observed in all cases.

Image analysis and droplet tracking. To quantify the dynamics of chasing droplets, we identified droplet positions using the Matlab imfindcircle algorithm, which is based on a Hough transformation. Subsequent tracking of the positions was performed by using the 2007 Matlab implementation by Blair and Dufresne of the Crocker and Grier tracking software(*30*) which was downloaded from http://site.physics.georgetown.edu/matlab/. Matlab version R2018B was used.

**Supplementary Text**

Quantification of interaction energy and displacement energy for two-body droplet chasing. To measure the energies associated with the predator-prey interaction of a droplet pair, we measured both the change in the droplet separation distance with time, $\delta r / \delta t$, and the velocity of the midpoint between the droplets, $v_{midpoint}$, as a function of the inter-particle distance, $r$. At droplet separations larger than 130 μm, the droplet velocities due to solute-mediated interactions were small compared to the drift velocity, so we only measured interactions for droplets closer than 130 μm and considered droplets at larger separations to be isolated. To minimize contributions from drift in the calculations, we measured $v_{midpoint}$ and $\delta r/\delta t$ for isolated droplets (which do move from drift) and subtracted those values from velocity measurements for interacting droplets at small separation distances so that we isolate the component of the droplet speed that is due to chemotactic interactions. We averaged $v_{midpoint}$ and $\delta r/\delta t$ as a function of drop separation distance attained from 3 to 5 separate experiments using bins of 1 μm in inter-droplet separation.

At low Reynolds numbers, the energy associated with the interaction between two droplets $E_{interaction}$ equals the area under the curve $\delta r/\delta t$ versus $r$ times the drag constant $C_D$, where we took the drag constant to be equal to the Stokes' drag multiplied by a dimensionless factor $C$ that corrects for flow inside the droplet and proximity of the droplet substrate. We thus found $\frac{E_{interaction}}{C} = 6\pi\eta a \int_{r=2a}^{r=\infty} \delta r/\delta t \, dr$, where $\eta = 0.89$ mPa s is the viscosity of the aqueous solution and $a$ is the droplet radius. The sign of this interaction energy does not indicate whether energy was consumed or produced, but rather whether the interaction was repulsive (positive) or attractive (negative). The energy associated with the displacement of the midpoint of the chasing pair is $\frac{E_{displacment}}{C} = 6\pi\eta a \int_{r=2a}^{r=\infty} v_{midpoint} \, dr$. The sign of the displacement energy indicates the direction of the chase, where we arbitrarily defined a positive displacement energy to correspond to a chase in which the fluorinated oil is prey (as seen in Figure 2).

Quantification of interaction parameters for simulations. We compared experimentally observed cluster dynamics of bromooctane (BOct) droplets mixed with ethoxynonafluorobutane (EFB) or methoxynonafluorobutane (MFB) droplets with overdamped particle dynamics simulations of particles interacting through chemotactic forces. We used the form of the speed of droplet 1 due to chemotactic forces exerted by droplet 2: $\overrightarrow{v_{12}} = C_{12} \frac{\overrightarrow{r_{12}}}{|\overrightarrow{r_{12}}|^3}$ that was predicted by Golestanian(17), where $\overrightarrow{r_{12}}$ is the center-to-center separation between drop 1 and 2 and $C_{12}$ is an interaction constant with units μm$^3$/s. To compare the multibody simulations directly with experiments, we set out to extract from observed two-body droplet interactions the interaction constants, $C_{qp}$ and $C_{pq}$, where we introduce the convention that "p" indicates the predator (BOct) and "q" indicates the prey (EFB or MFB). $C_{qp}$ determines the speed $\overrightarrow{v_{qp}}$ with which a prey droplet moves due to the chemotactic interaction with a predator, and $C_{pq}$ similarly determines the predator speed $\overrightarrow{v_{pq}}$ due to the interaction with a prey droplet. To this end, we tracked droplet positions during two-droplet encounters between a predator and a prey and measured the component of the velocity of the prey that points away from the midpoint between the prey and predator: $v_{qp} = \frac{\overrightarrow{v_q} \cdot (\overrightarrow{r_q} - \overrightarrow{r_p})}{|\overrightarrow{r_q} - \overrightarrow{r_p}|}$, where $\overrightarrow{r_q}$ and $\overrightarrow{r_p}$ are the position of the prey and predator droplet respectively. The speed $v_{qp}$ is positive if the prey moves away from the predator and negative if the prey moves toward the predator. **Figure S4** shows that a graph of $v_{qp}$ versus the inverse squared droplet separation $r_{qp}^{-2}$ is approximately a straight line from which we extracted the interaction constant $C_{qp}$ using the Matlab least squares fitting procedure. Similarly, we measured the interaction constant $C_{pq}$ from a curve of the experimentally measured $v_{pq}$ versus $r_{qp}^{-2}$.

We did not observe interactions of sufficient strength between two prey drops to measure $C_{qq}$, which determines the speed of prey due to interactions with another prey, or between two predator droplets to measure $C_{pp}$, which determines the speed of a predator due to interaction with another predator. To find those two interaction parameters $C_{qq}$ and $C_{pp}$, we measured the distance between two predator drops, $r_{pp}$, in a cluster of two predators and one prey as shown in **Figure S5**. This distance $r_{pp}$ is set by a balance of the component of the speeds of the predators toward each other due to their interaction with the prey and the speeds of the predators away from each other due to mutual predator-predator repulsion:

$$\frac{C_{pp}}{r_{pp}^2} = \frac{C_{qp}\sin(\frac{1}{2}\theta)}{r_{qp}^2}. \qquad \text{Eq. S1}$$

Here $\theta$ is the angle the two predators and the prey make with each other, $r_{pp}$ is the center-to-center distance between the predators, and $r_{qp}$ is the center-to-center distance between the predator and prey, which at contact is given by the sum of predator and prey radii. The distances $r_{qp}$ and $r_{pp}$ were measured from the cluster geometry and $C_{qp}$ was independently measured from two-body chasing interactions as just previously described, so that we can calculate $C_{pp}$. Finally, we find $C_{qq}$ using the fact that the four interaction constants $C_{qq}, C_{qp}, C_{pq}$ and $C_{pp}$ are related through:

$$C_{qq} = \frac{C_{qp}C_{pq}}{C_{pp}}. \qquad \text{Eq. S2}$$

The relation in Equation S2 uses the fact that the interaction constant $C_{12} = \frac{\mu_1 \alpha_2}{D}$ depends on the mobility of droplet 1, $\mu_1$, which determines the speed of the droplet in a solute gradient due to the Marangoni effect and on the activity $\alpha_2$, that represents the rate at which molecules are produced or consumed at the surface of droplet 2(*17*). This activity divided by the diffusion constant of the solute, D, determines the solute gradient that causes droplet 1 to move. Using this definition we find that $\frac{C_{qp}C_{pq}}{C_{pp}} = \frac{\mu_q \alpha_p \mu_p \alpha_q D}{\mu_p \alpha_p D^2} = \frac{\mu_q \alpha_q}{D} = C_{qq}$. We list the interaction values we found for interactions between BOct and EFB or MFB droplets in **Table S1**.

We initialized the simulations of multi-body droplet interactions by placing $n_p$ predator and $n_q$ prey drops on positions that match the positions in an experiment. Then, we calculated the speed $\vec{v_p}$ of each predator drop $p$ and prey drop $q$, $\vec{v_q}$, by summing the contributions of each other droplet to the phoretic velocity as

$$\vec{v_p} = \sum_{i=1, i \neq p}^{n_p} C_{pp} \frac{\vec{r_{ip}}}{|\vec{r_{ip}}|^3} + \sum_{j=1}^{n_q} C_{pq} \frac{\vec{r_{jp}}}{|\vec{r_{jp}}|^3}, \qquad \text{Eq. S3}$$

$$\vec{v_q} = \sum_{i=1}^{n_p} C_{qp} \frac{\vec{r_{iq}}}{|\vec{r_{iq}}|^3} + \sum_{j=1, j \neq q}^{n_q} C_{qq} \frac{\vec{r_{jq}}}{|\vec{r_{jq}}|^3}, \qquad \text{Eq. S4}$$

Here $i$ and $j$ are indices that designate predator and prey drops respectively. The displacement in a time step $dt$ is then given by $d\vec{r} = \vec{v}dt$. We chose the time step $dt$ such that it was always smaller than the smallest drop radius divided by the largest possible speed for a given set of initial conditions. Typically $10^{-2}$ s $< dt < 10^{-3}$ s. We chose the number of time steps $n$ based on the length of the experimentally observed interaction and *n* typically varied from $n = 10^3$ to n=$10^4$. We treated the droplets as hard spheres and resolve overlap using inelastic collisions.

Because we measured three of the four interaction constants directly, $C_{pp}, C_{qp}, C_{pq}$ and calculated the fourth constant $C_{qq}$ from Equation S2, we can directly compare our simulations with the experiments. We tuned the parameters used in the simulations within the uncertainty range of the measured values to match the experimentally observed dynamics as closely as possible.

Trochoidal trajectories. Based strictly on the concept of chemotactic forces, we would expect that a predator droplet chasing after prey would always move in a straight line, assuming no other forces from neighboring droplets are introduced. Instead, we commonly observed trochoidal trajectories with a steady pitch for BOct chasing of either EFB or MFB, as shown in **Figure 4B**. Such trajectories were most often observed when the BOct droplets were small, under approximately 40 μm diameter. Based on the observation that BOct droplets become self-propelled and move at a significantly faster velocity when they shrink to a diameter below about 40 μm (**Figure S3**), we speculate that the reason for this trochoidal motion is that the BOct predator drop moves along the prey's surface during the chase due to its self-propulsion. This rearrangement of the predator on the surface of the prey causes the prey droplet to turn in response, because it always moved away linearly from the predator, leading to a redirect of the chase direction.

**Figure S6** shows the absolute speed of a predator and prey droplet in a typical trochoidal chase. The predator drop indeed moved slightly faster than the prey; however, this phenomenon should cause the pair to swim in circles, not along trochoidal trajectories. However, if there is any drift in the system, which we often observe presumably due to large-scale solute gradients, the circular trajectories turn into trochoidal trajectories. In agreement with that idea, the speed of both the predator and prey particle oscillated and was higher when moving in the direction of drift.

Efficiency of chasing driven by oil exchange. We consider here the motion of a pair of droplets that exchange oil with one another through the aqueous Triton X-100 surfactant phase, where the two droplets consist of iodo-*n*-alkanes of differing *n*. The iodoalkanes are miscible in any ratio so that, when the droplets are in close proximity, there is iodoalkane transport from the droplet through the surfactant solution phase into the neighboring droplet and vice-versa. For droplets of different iodoalkanes, this oil exchange is asymmetric such that there is a net oil transport between droplets. The asymmetric oil transport drives chasing and propels the motion of the droplet pair with speeds of 20 μm/s or more through coupling with the local droplet interfacial tensions and the Marangoni effect.

To estimate the efficiency of this propulsion mechanism, we assume that the mixing of the two oils is ideal and ignore contributions from the surface energy. The assumption that the mixing is ideal is reasonable because the two oils are miscible in any ratio and are chemically very similar. Any enthalpic effects that are ignored would most likely only decrease Gibbs free energy of mixing, rendering this estimate of the propulsion efficiency a lower boundary. For droplets of similar diameter, any contributions of the surface energy to the Gibbs free energy change of oil exchange would be small. To calculate the efficiency, we compare the Gibbs free energy of mixing with the displacement work of the pair of droplets. To simplify the calculation, we consider specifically the input power (i.e. the Gibbs free energy of mixing per unit time) with the output power, which is related to the velocity of the droplets.

We make an order of magnitude estimate of the output power $P_{out}$ from the velocity of the chasing droplet pair, assuming that the drag on these droplets is the Stokes' drag and ignoring modifications due to the proximity of the wall, nearby droplets, and the fact that the drops are fluid; thus, the force exerted on a droplet pair, $F_{pair}$, causing it to move with a velocity $v_{pair}$ is $F_{pair} \approx 6 \pi \eta a\, v_{pair}$, where $\eta = 0.89$ mPa·s is the viscosity of the aqueous surfactant solution and $a$ is the radius of the droplets. The power consumption by the pair of droplets is then simply the force exerted on the droplets multiplied by the distance over which the force was applied and divided by the time it took to cross that distance, so we find

$$P_{out} \approx 6 \pi \eta a\, v_{pair}^2 \qquad \text{Eq. S5}$$

For example, for droplets with a diameter of 70 μm that move at 20 μm/s, the dissipated power is $P_{out} \approx 2.3 \times 10^{-16}$ J/s. This power required to move the pair of droplets comes from a change in the Gibbs free energy of mixing of the two oils, $\Delta_{mix}G = -T\Delta_{mix}S$, where we assume ideal mixing. For a binary mixture, the entropy gain is $\Delta_{mix}S = -n R\, (x \ln(x) + (x - 1)\ln(x - 1))$, where $n$ represents the amount of mixed oil in moles, $R = 8.314 \frac{\text{J}}{\text{mol K}}$ is the ideal gas constant, and $x$ is the mole fraction of one oil in the other after mixing. The maximum energy gain corresponds to mixing of equal molar amounts of oil in a 1:1 mixture, where $x = 0.5$ and the entropy of mixing simplifies to $\Delta_{mix}S = -n R \ln(2)$ such that the Gibbs free energy of mixing is

$$\Delta_{mix}G = -n R T \ln(2) \qquad \text{Eq. S6}$$

Equation S6 gives an estimate of the Gibbs free energy change associated with the transition from pure droplets of two different oils to equimolar mixed oil droplets. The Gibbs free energy of this exchange divided by a typical timescale over which the exchange occurs, $\Delta t$, gives the power production available to move the droplet pair. Assuming that the slowest step in the oil exchange process is transport into the aqueous phase, we estimate the oil exchange time from the measured solubilization rates of individual droplets, given in **Figure 3B**. We estimate that the oil exchange time is the time in which the volume of one droplet can be replaced entirely due to the combined transport rates of both droplets: $\Delta t \approx V_0 \left(\frac{dV_1}{dt} + \frac{dV_2}{dt}\right)^{-1}$, where $V_0$ is the droplet's initial volume and $\frac{dV_1}{dt}$ and $\frac{dV_2}{dt}$ are the measured solubilization rates of both droplets. The average power produced by oil mixing ($P_{in}$) is the Gibbs free energy change of mixing divided by the mixing time, $P_{in} = \Delta_{mix}G/\Delta t$. Substituting for $\Delta_{mix}G$ using Eq. S6 and substituting for $\Delta t$, we find

$$P_{in} = -\frac{1}{V_0}\left(\frac{dV_1}{dt}n_1 + \frac{dV_2}{dt}n_2\right) R T \ln(2) \qquad \text{Eq. S7}$$

As an example, we measured that a pair consisting of one iodononane droplet and one iodopentane droplet with diameters of approximately 70 μm moves at 18 μm/s, so that the power required to move the pair is approximately $P_{out} = 1.9 \times 10^{-16}$ J/s. Individually, the iodononane and iodopentane droplets solubilize at rates of $\frac{da}{dt} = 3.05$ nm/s and $\frac{da}{dt} = 3.85$ nm/s, respectively. We use fact that the droplet size does not change significantly over the timescale on which chasing occurs such that $\frac{dV}{dt} = 4\pi a^2 \frac{da}{dt}$, and find $n$ for each iodoalkane individually by dividing the droplet volume over the molecular volume of the oil $v_m = \frac{M_w}{\rho}$ where $M_w$ is the oil molecular weight and $\rho$ the density. We estimate a power production $P_{in} = -1.2 \times 10^{-9}$ J/s using Eq. S7. Comparing $P_{out}$ and $P_{in}$ we find an efficiency $\epsilon = 1.6 \times 10^{-7}$. In conclusion, we find that the lower boundary of the swimming efficiency of the pairs of droplets described in this report is on the order of $10^{-7}$ which is two orders of magnitude higher than the typical efficiencies of $10^{-9}$ found for colloidal Janus swimmers whose motion is driven by the decomposition of hydrogen peroxide(*21*).

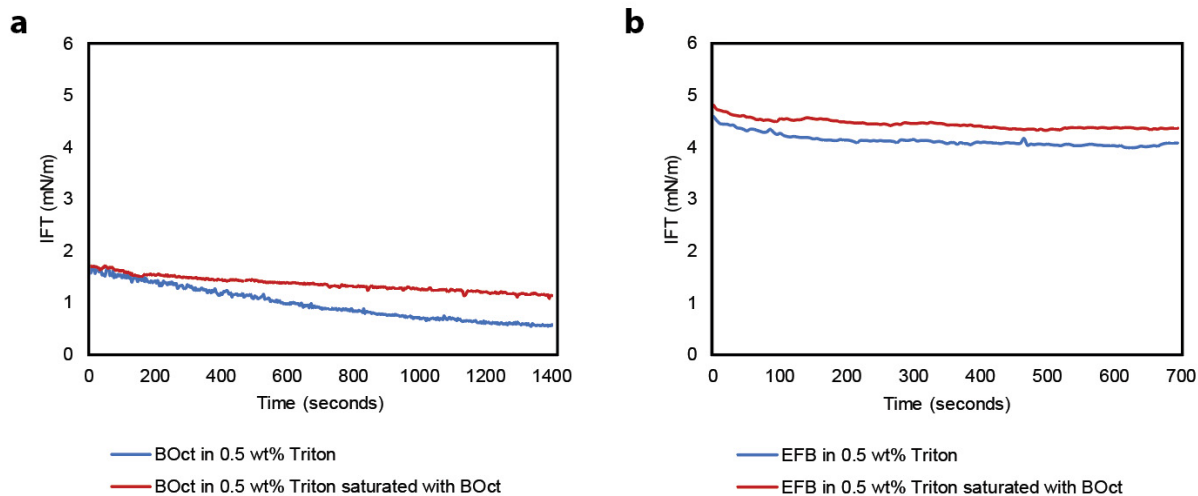

**Figure S1. The pendant drop method is used to measure the dynamic interfacial tension (IFT) increase resultant from BOct oil saturation in Triton surfactant solutions. a,** The interfacial tension of BOct is higher in BOct-saturated 0.5 wt% Triton surfactant solution (red) compared to unsaturated surfactant solution (blue). **b,** A higher interfacial tension was also observed for EFB in BOct-saturated 0.5 wt% Triton (red), compared to oil-free Triton (blue). Experiments were produced in triplicate and the same trends were consistently observed.

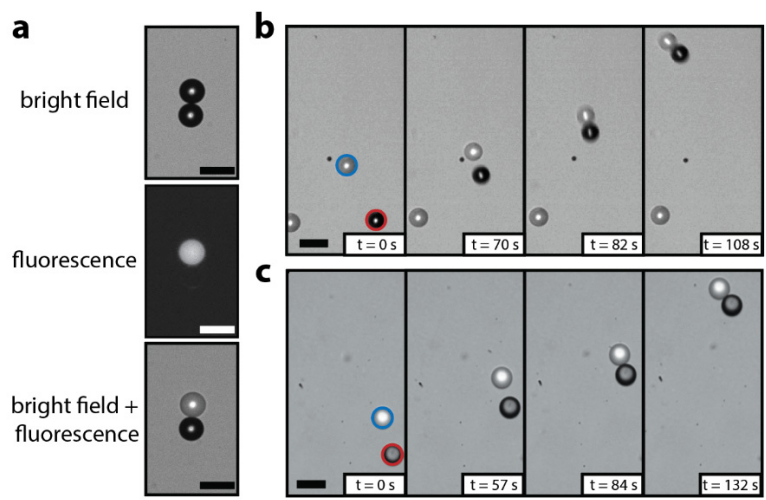

**Figure S2. Fluorescence was used to distinguish between chasing iodoalkanes or chasing alkanes and identify predator and prey. a**, Under transmission brightfield imaging, iodohexane and iodoheptane droplets are indistinguishable (top, and **Video S4**); by dying the iodohexane droplet with fluorescent Lumogen F Red 305, the iodohexane droplet can be easily identified using fluorescence microscopy (middle). By simultaneously imaging the droplets in both transmission brightfield and fluorescence (bottom), we can visualize all droplets and also distinguish the two types of oil. This imaging method is used in (b,c). **b**, An iodoheptane droplet (outlined in red) chases an iodohexane droplet dyed with fluorescent Lumogen F Red 305 (outlined in blue) in 0.5 wt% Triton. The optical micrographs were taken with simultaneous transmission brightfield and fluorescence imaging as shown in (a). Scale, 100 μm. **c**, An octane droplet (outlined in red) chases a hexadecane droplet dyed with fluorescent Lumogen F Red 305 (outlined in blue) in 0.5 wt% Triton. Optical micrographs were taken with simultaneous transmission brightfield and fluorescence imaging as shown in (a). Scale, 100 μm.

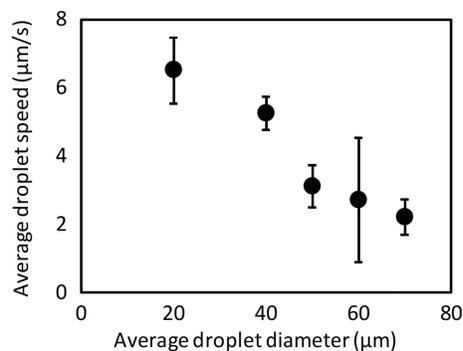

**Figure S3. BOct droplets become self-propelled at smaller diameters.** Droplet-tracking image analysis was used to determine the average droplet speed of BOct droplets as a function of droplet diameter as the drops were solubilizing in 0.5 wt% Triton. The average speed was calculated by measuring the nominal speed and subtracting the component of the speed that all droplets had in the same direction due to drift. Each data point represents between 20 and 50 droplet measurements, and the error bars represent the standard deviation. As the droplets become smaller, the average self-propulsion speed of the BOct droplets increases.

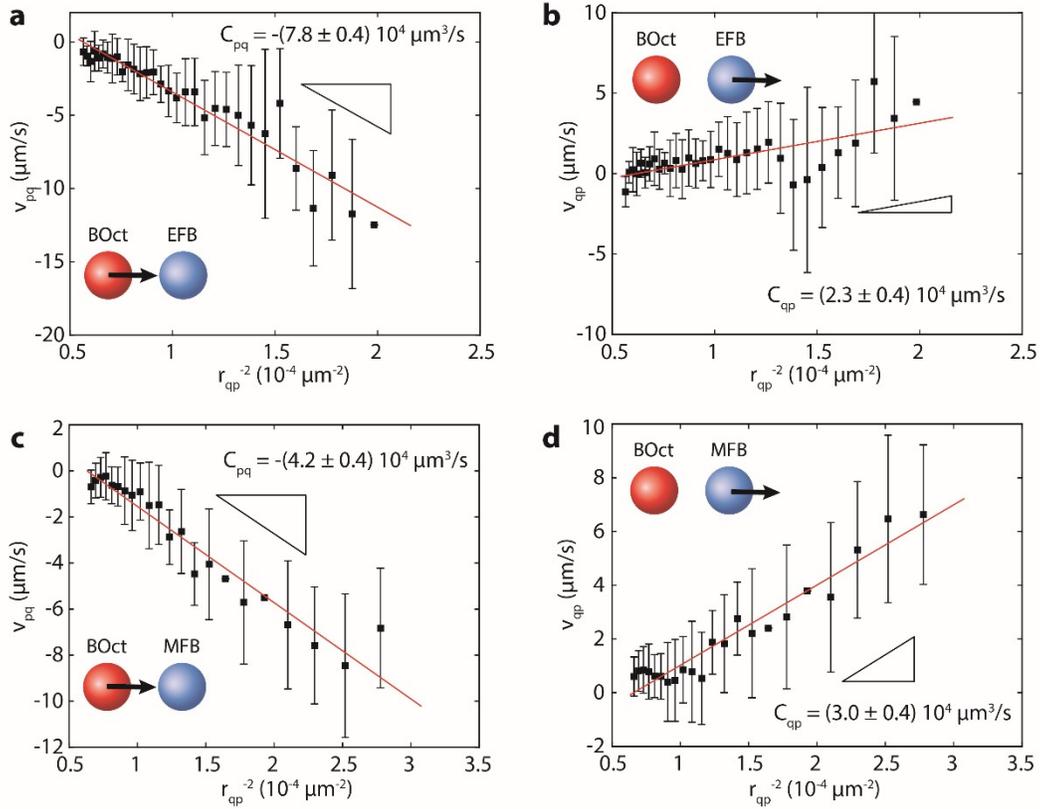

**Figure S4. Measurements of the two-droplet interaction constants $C_{qp}$ and $C_{pq}$ for encounters between BOct drops and either EFB or MFB droplets**. **a**, Approach speed of BOct droplets near EFB droplets. **b**, Escape speed of EFB droplets near BOct droplets. **c**, Approach speed of BOct droplets near MFB droplets. **d**, Escape speed of MFB droplets near BOct droplets. The data in each plot are measured from three separate two-droplet encounters. The speed versus distance curves were binned with 1 μm bins. The data represent the average speed per bin and the error bar represents the standard deviation.

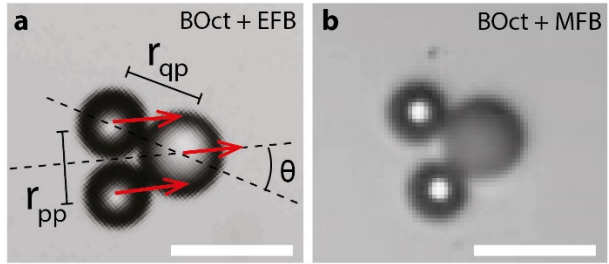

**Figure S5. Spacing between predator droplets when two predators chase a single prey. a**, Two BOct droplets chasing a single EFB droplet. The BOct droplets nearly touch. **b**, Two BOct droplets chasing an MFB droplet. The spacing between two BOct droplets is much larger than in a similar cluster with an EFB droplet. Scale bars are 100 μm.

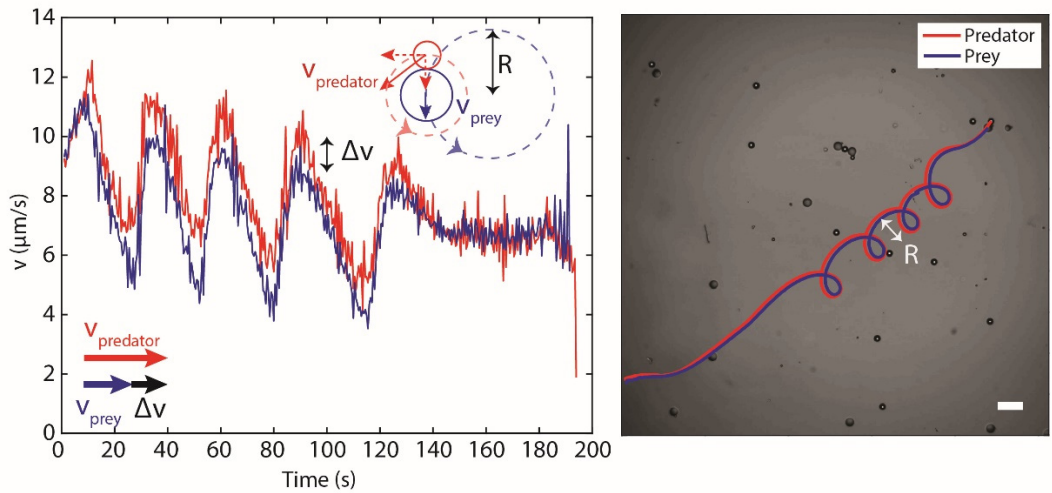

**Figure S6. Trochoidal motion of a swimming predator-prey pair. a**, Speed of the predator BOct droplet (red) and prey MFB droplet (blue) as function of time. The schematic indicates how self-propulsion of the predator drop along the surface of the prey results in a circular trajectory of the chasing pair. **b**, Trochoidal trajectory of a BOct droplet chasing after an MFB droplet. Scale, 100 μm.

**Table S1. List of experimentally determined interaction parameter values obtained from two-body chasing encounters of BOct droplets with EFB or MFB droplets as well as the specific value of each parameter used in the simulations.** Results of the simulations with the given parameters are shown in **Figure 4**. The parameters used in the simulations were chosen from within the experimentally determined range to yield droplet dynamics that most closely matched the experiment.

|  | BOct and EFB | | BOct and MFB | |
| --- | --- | --- | --- | --- |
|  | Experiment | Simulation | Experiment | Simulation |
| $C_{pp}$ ($10^4$ µm$^3$/s) | -4.8 ± 1.6 | -4.8 | -4.8 ± 1.6 | -4.9 |
| $C_{pq}$ ($10^4$ µm$^3$/s) | 2.3 ± 0.4 | 2.3 | 2.9 ± 0.4 | 2.5 |
| $C_{qp}$ ($10^4$ µm$^3$/s) | -7.8 ± 0.4 | -7.8 | -4.2 ± 0.4 | -3.8 |
| $C_{qq}$ ($10^4$ µm$^3$/s) | 3.7 ± 0.4 | 3.7 | 3.2 ± 1.0 | 4.2 |

**Video S1. A bromooctane droplet chases an ethoxynonafluorobutane droplet in 0.5 wt% Triton X-100 surfactant.** The bromooctane predator droplet (darker colored drop) accelerates as it moves towards the ethoxynonafluorobutane prey (lighter colored drop), and the prey droplet flees but is eventually "caught". The resulting droplet pair continues to move in a direction always led by the ethoxynonafluorobutane prey droplet. Still frames from this video are included in Figure 1a. Scale bar, 100 μm. Video speed, 10x real time.

**Video S2. Nonreciprocal chasing interactions produce droplet pairs and mixed multibody droplet clusters.** Chasing pairs of bromooctane (darker colored drops) and ethoxynonafluorobutane (lighter colored drops) and larger multi-droplet clusters undergo various motions as a result of nonreciprocal interactions between droplets. The continuous phase is 0.5 wt% Triton X-100 surfactant solution. Scale bar, 250 μm. Video speed, 25x real time.

**Video S3. Ethoxynonafluorobutane droplets chase bromooctane droplets in 3 wt% Capstone FS-30 surfactant.** Ethoxynonafluorobutane predator droplets (lighter colored drops) chase bromooctane prey droplets (darker colored drops) in 3.0 wt% fluorosurfactant Capstone FS-30. This chasing direction is the reverse of that observed in 0.5% Triton X-100 (as seen in **Videos S1** and **S2**) due to the fact the fluorosurfactant preferentially solubilizes and transports the fluorinated oil. Scale bar, 250 μm. Video speed, 25x real time.

**Video S4. 1-Iodoheptane droplets chase 1-iodohexane droplets in 0.5 wt% Triton X-100 surfactant.** The designation of chasing direction was achieved by fluorescent labeling (Extended Data Figure 2) Scale bar, 250 μm. Video speed, 25x real time.

**Video S5. Marangoni-driven directional flows inside the predator and prey droplets align with the droplets' direction of motion.** A 1-iodoheptane droplet chases a 1-iodohexane droplet in 0.5% wt% Triton X-100, and the droplet-internal flows are visualized with iron oxide tracer particles suspended in both oils. Scale bar, 50 μm. Video speed, 1x real time.

**Video S6. Collective assembly and disassembly dynamics of droplet clusters.** A cascade of nonreciprocal interactions between bromooctane (darker drops) and ethoxynonafluorobutane (lighter drops) in 0.5 wt% Triton X-100 result in dynamic behavior of larger droplet clusters. Exchange of droplets between different clusters is commonly observed. Scale bar, 500 μm. Video speed, 25x real time.

**Video S7. Dynamic cluster motions resultant from self-propelled droplets.** Bromooctane predators (darker colored droplets) with diameters below approximately 40 μm in 0.5 wt% Triton can become self-propelled, resulting in trochoidal, run-and-tumble, and flapping motions. Scale bar, 250 μm. Video speed, 50x real time.

**Video S8. Simulations reproduce experimentally observed cluster dynamics.** The circular motion of a droplet cluster containing two bromooctane drops (darker, red colored) and two ethoxynonafluorobutane drops (lighter, blue colored) was reproduced in a simulation by using droplet initial positions and measured two-body droplet interaction parameters as inputs. Continuous phase is of 0.5 wt% Triton X-100 surfactant solution. Scale bar, 100 μm. Video speed, 10x real time.

**Video S9. Individual droplets act as chemical signaling posts to direct the motion of chasing droplet pairs as seen in experiment and simulation.** Seen in both experiment and simulation, individual bromooctane droplets (darker, red drops) act as chemical signaling posts via long-range solute-mediated interactions to direct the movement of a bromooctane and ethoxynonafluorobutane chasing droplet pair. Continuous phase of 0.5 wt% Triton X-100 surfactant solution. Scale bar, 100 μm. Video speed, 10x real time.

**Video S10. Solute-mediated interactions between neighboring droplets cause cluster rearrangement as seen in experiment and simulation.** Repulsive interactions between bromooctane predators (darker, red drops) chasing ethoxynonafluorobutane prey (lighter, blue drops) in separate clusters results in the destabilization and loss of a single bromooctane droplet from one of the clusters. Continuous phase of 0.5 wt% Triton X-100 surfactant solution. Scale bar, 100 μm. Video speed, 10x real time.